\documentclass[aps,amsmath,amssymb,amsfonts,superscriptaddress,nofootinbib]{revtex4}
\usepackage[english]{babel}

\newcommand{\bv}{{\bf v}}
\newcommand{\br}{{\bf r}}

\newcommand{\beq}{\begin{equation}}
\newcommand{\eeq}{\end{equation}}
\newcommand{\bea}{\begin{eqnarray}}
\newcommand{\eea}{\end{eqnarray}}

\newcommand{\ga}{\mbox{${\gamma}$}}

\newcommand{\De}{\mbox{${\Delta}$}}

\newcommand{\om}{\mbox{${\omega}$}}
\newcommand{\Om}{\mbox{${\Omega}$}}

\newcommand{\gsim}{\mbox{$>$\hspace{-0.8em}\raisebox{-0.4em}{$\sim$}}}
\newcommand{\lsim}{\mbox{$<$\hspace{-0.8em}\raisebox{-0.4em}{$\sim$}}}

\begin{document}

\title{Passage of small black hole through the Earth.\\ Is it detectable?}

\author{\firstname{I.~B.}~\surname{Khriplovich}}
\email{khriplovich@inp.nsk.su}
\author{\firstname{A.~A.}~\surname{Pomeransky}}
\email{a.a.pomeransky@inp.nsk.su}
 \affiliation{Budker Institute of
Nuclear Physics, 630090, Novosibirsk, Russia, and Novosibirsk
University}
\author{\firstname{N.}~\surname{Produit}}
\email{Nicolas.Produit@obs.unige.ch} \affiliation{INTEGRAL Science
Data Center, 16, Chemin d'Ecogia, CH-1290 Versoix, Switzerland}
\author{\firstname{G.~Yu.}~\surname{Ruban}}
\email{gennady-ru@ngs.ru} \affiliation{Budker Institute of Nuclear
Physics, 630090, Novosibirsk, Russia, and Novosibirsk University}

\begin{abstract}
We examine the energy losses of a small black hole passing through
the Earth, and in particular, the excitations created in the
frequency range accessible to modern acoustic detectors. The
dominating contributions to the effect are due to the coherent
sound radiation of the Cherenkov type and to the conversion of
black hole radiation into sound waves.
\end{abstract}

\maketitle

\section{Introduction}

Black hole is a body with an extremely strong gravitational
attraction. Any object, if its distance from this body is smaller
than some critical one (so-called gravitational radius $r_g$),
cannot overcome this attraction: such an object moves to the black
hole, but cannot move away from it. In other words, a black hole
swallows everything sufficiently close to it (hence its name). The
existence of black holes is predicted by general relativity.

Heavy black holes are an inevitable result of the evolution,
compression of sufficiently heavy stars. There are direct
observational evidences of their existence, in particular in some
binary stars. One component of such a binary has a mass $M \gsim
3M_\odot$ ($M_\odot$ is mass of the Sun). This component is a
strong X-ray source. The only reasonable explanation of the
observed effects is the gas accretion from the normal component to
this one, which is a black hole.

Besides, detailed investigations
of star motion in some galaxies (including ours) indicate that
their nuclei contain superheavy black holes of mass $M \sim
4\times 10^7 M_\odot$.

It can be demonstrated, however, that if the mass of a star is
$\,\lsim \,2M_\odot$, its gravitational field is insufficient to
compress this star (with the radius of about $10^6$ km) to its
gravitational radius $r_g$, which is of about few kilometers only.
So, can light black holes exist?

Such black holes could arise at the early stages of the Universe
evolution (this is why they are called primordial) when the matter
density was very high, from inhomogeneities of this density. But
which of these small black holes could survive since those times?
The problem is their thermal radiation, a subtle
quantum-mechanical effect. Its intensity is proportional to
$1/M^2$. Thus, too light primordial black holes have already
evaporated and disappeared due to this radiation. Estimates
demonstrate that the masses of the survivors should not be less
than $5 \times 10^{14}$ g. By common standards, such a mass of
such a survivor is huge, but its size is tiny, $r_g \sim 10^{-12}$
cm.

These objects have never been observed, though their existence
does not contradict any known law of nature. As a step to study
the possibility to detect the passage of such an object through
the Earth (or some other planet, or the Moon), we discuss here the
effects arising during this passage. Detailed account of the
calculations, as well as the analysis of some other possible effects
(perhaps, too tiny to be of importance for the problem), is
presented elsewhere~\cite{kp}.

\section{Coherent sound generation by a supersonic black hole}

We start with the dynamics of mechanical deformations and
excitation of sound waves caused by the gravitational field of a
primordial black hole passing through the Earth. It will be
assumed that the velocity $v$ of the black hole is comparable to
that of the Earth, which exceeds the speed of sound in the matter
$c_s$.

Then the elastic gravitational scattering of a black hole in the matter
results in supersonic sound radiation of the Cherenkov type with
the total intensity
\beq\label{iel}
 I_{\rm el}= 4\pi(G M)^2 \rho/v\,\ln\frac{k_1 c_s}{\omega_p}\,.
 \eeq
Here $G$ is the Newton gravitational constant, $M$ is the mass of
black hole, $\rho$ is the density of matter, $\om_p = (4\pi G
\rho)^{1/2}$, $k_1$ is the maximum momentum transfer at which the
scattering remains elastic; we assume that $v \gg c_s $.

In the case of large momentum transfers, $k > k_1$, the matter can
be considered as a collection of free particles. Then the
scattering of the black hole reduces to that on a free particle.
Thus obtained rate of inelastic energy loss by a black hole is
\beq\label{in}
I_{\rm inel} = 4\pi(G M)^2 \rho/v\, \ln\frac{1}{k_1 a}\,,
\eeq
where $a$ is the typical interatomic distance in the matter.

Finally, the total rate of energy loss is
\beq\label{tot}
I_{\rm tot} = I_{\rm el} + I_{\rm inel} = 4\pi(G M)^2 \rho/v\,
\ln\frac{c_s}{\omega_p\, a}\,.
\eeq
With the accepted logarithmic accuracy, this total rate is
independent of the critical momentum transfer $k_1$. On the other
hand, for any reasonable choice of $k_1$, the elastic energy loss
dominates strongly, $I_{\rm el} \gg I_{\rm inel}$. It is worth mentioning
also that the logarithm in (\ref{tot}) is really large, about 35.
An expression for $I_{tot}$ close to (\ref{tot}), but with a
different logarithmic factor, was obtained previously by K.
Penanen \cite{Penanen}.

To estimate the energy $\De E$ released by a black hole passing
through the Earth, this rate should be multiplied by the time of
the passage, $\tau = L/v$. For numerical estimates we assume that
the equilibrium density of matter is $\rho = 6$~g/cm$^3$, the path
$L$ is about the Earth diameter, $L \sim 10^4$ km, and the
velocity of black hole is $v \sim 30$ km/s. Then, for a black
hole with mass $M  \sim 10^{15}$ g this energy loss constitutes
about
\beq\label{rel}
\De E \sim 4\times 10^{9}\; {\rm J}\,.
\eeq

Let us note that this energy is much smaller than that released at
the explosion of a 10 kiloton atomic bomb
\beq\label{bomb}
\De E_{\rm bomb} \sim 5 \times 10^{13}\; {\rm J}\,.
\eeq
Besides, when comparing the energy released by a black hole (not
only (\ref{rel}), but also other contributions to it considered
below) with the energy of an atomic bomb, one should keep in mind
that the source of $\De E_{\rm bomb}$ is practically point-like,
while $\De E$ is spread along a path $L~\sim~10^4$~km. Of course,
the energy released by a black hole is extended not only in space,
but in time: it takes several minutes for the black hole to cross
the Earth, but the release of energy in an atomic bomb or in an
earthquake happens in much shorter time intervals. Of course, due
to this extension of the effect both in space and time, it is much
more difficult to detect the passage of a small black hole.

To discover a mini-black hole passing through the Earth, one has
to study seismic vibrations induced by this passage. The
sensitivity of appropriate seismic detectors is confined to the
frequencies in the interval around $\om_{min} \sim 0.1$ Hz and
$\om_{max} \sim 10$ Hz. As to the frequency distribution of the
acoustic Cherenkov radiation, it can be demonstrated that with $v
\gg c_s$ and $\omega \gg \omega_p$, the result is
\beq
dI_{\rm el}\,=\,4\pi(GM)^2(\rho/v)\,\frac{d\om}{\om}\,.
\eeq
Thus, the energy of the vibrations excited in the frequency
interval $\om_{min} \div \,\om_{max}$ is
\beq
\De E^{\omega} = 4\pi(G M)^2 L\rho/v^2\,
\ln\frac{\om_{max}}{\om_{min}}\,.
\eeq
For the discussed frequency interval, 0.1 $\div$ 10 Hz, it
constitutes numerically
\beq\label{relom}
\De E^{\omega} \sim \,5 \times 10^{8}\; {\rm J}\,,
\eeq
or about 1/10 of the total energy (\ref{rel}).

Let us note that similar results hold for the energy losses of a
mini-black hole in water, which could be of interest for the
underwater cosmic-ray acoustic detectors. Of course, the
underwater path $L$ is much less than the Earth diameter, and the
density of the medium changes here from $\rho \simeq  6$ g/cm$^3$
to $\rho = 1$ g/cm$^3$. Though the typical frequencies propagating
in water, both $\om_{min}$ and $\om_{max}$, are much higher than
those in the Earth, the value of $\ln\om_{max}/\om_{min}$ can be
considered to be about the same to our accuracy.

\section{Conversion of black hole radiation into sound waves}

One more source of the energy transfer from a light black hole to
the matter (though not its kinetic energy discussed above, but the
internal one) is the black hole radiation. Of course, for our
purpose we have to consider the emission of $\ga$ and $e^\pm$ only
(but not gravitons and neutrinos). Using the results of
\cite{Page}, we obtain under the same assumptions ($M  \sim
10^{15}$ g, $L \sim 10^4$~km, and $v \sim 30$ km/s) the following
estimate for the total radiation loss of such black hole:
\beq\label{rad}
\De E_{\rm rad} \sim 1.5 \times 10^{12}\; {\rm J}\,.
\eeq

The leading mechanism for the conversion of this radiation into
sound waves is as follows. The radiation absorbed by matter
increases the temperature along the path of the source. It results
in the inhomogeneous and non-stationary thermal expansion of the
matter and thus in the emission of acoustic waves. This effect
permits of rather reliable theoretical analysis, which leads to
the following equation for pressure~\cite{wes}:
\beq \label{source}
\frac{1}{c_s^2}\ddot{p}-\Delta p = \frac{\beta}{C} \dot{W}\,.
\eeq
Here $\beta= - 1/\rho\; (\partial \rho/\partial T)_p$ is the
coefficient of thermal expansion, $C$ is the specific heat, and
$W$ is the power density. A black hole can be treated as a
point-like source of radiation with intensity $I$, so that in our
case $W=I \delta (\br - \bv t)$.

In this way one can calculate the mechanical energy acquired by
matter due to this thermal expansion. The result for the spectral
intensity of thus induced sound waves is
\beq
\frac{dE_m}{dt d\omega}=\left(\frac{\beta I}{C}\right)^2
\frac{\omega}{4\pi v \rho}\,.
\eeq
The estimate for the maximum value $\Om$ for the thus emitted frequencies
looks as follows:
\beq\label{Om}
\Om \sim \frac{c_s}{r_o}\,;
\eeq
here $r_0$ is the typical absorption length for $\ga$ and $e^\pm$ emitted by
the black hole. In the present case  $r_0$ is about 3 cm, so that $\Om$ is
much larger then the frequencies of interest.
Curiously, this sound radiation occurs if $v > c_s$. In other words,
this is also a sort of Cherenkov effect.

The total energy radiated at the frequency $\omega$,
during the passage of a black hole through the Earth, can be
conveniently written as
\beq\label{Em}
\frac{dE_m}{d\omega}=\left(\frac{\beta \Delta E_{\rm
rad}}{C}\right)^2 \frac{\omega}{4\pi L \rho}\,;
\eeq
we go over in this expression from the intensity $I$ of the black
hole radiation to the total energy $\Delta E_{\rm rad}$ emitted
during the passage through the Earth: $\Delta E_{\rm rad}=I\,L/v$.
The overall energy of the seismic waves generated by the black
hole radiation, released in the frequencies $\omega <
\omega_{\rm max}$ is, according to (\ref{Em}),
\beq
E^{\omega}_m =\left(\frac{\beta \Delta E_{\rm rad}}{C}\right)^2
\frac{\omega_{\rm max}^2}{8\pi L \rho}.
\eeq
With $C=1\,{\rm J}\, {\rm g}^{-1}\,{\rm K}^{-1}$, $\beta=0.5 \cdot
10^{-4}\, {\rm K}^{-1}$, and $\omega_{\rm max} = 10$ Hz, we obtain
$E^{\omega}_m \sim 1\,{\rm J}$. So, this effect is much less than
that of the Cherenkov sound radiation (\ref{relom}).

The situation changes essentially for the underwater effects. Here
the admissible frequencies extend to about $\omega_{\rm max} = 30$
kHz (which is already on the order of $\Om$), so that the
conversion of the black hole radiation into sound
waves gets about as effective as the Cherenkov sound radiation.

In conclusion, we wish to point out that though the effects of
radiation damage contribute negligibly into the seismic signal, they
can create quite a distinct pattern in crystalline material. The
dose deposited is estimated as \beq \frac{\De E_{\rm rad}}{\rho L
r_0^2}\; \sim \,10^{5}\; {\rm Gy} \quad \quad (\, 1 {\rm Gy}\; ({\rm
Gray}) = 1\;{\rm J/kg}\,)\,. \eeq It creates a long tube of heavily
radiatively damaged material, which should stay recognizable for
geological time.

\section{Acknowledgements}

We are grateful to N.M. Budnev, R. Mosetti and V.S. Seleznev for
useful discussions.


\begin{thebibliography}{99}
\bibitem{kp} I.B. Khriplovich, A.A. Pomeransky, N. Produit,
G.Yu. Ruban, Phys. Rev. {\bf D}, in press.
\bibitem{Penanen} K. Penanen, private communication.
\bibitem{Page} D.N. Page, Phys. Rev. {\bf D 13}, 198 (1976).
\bibitem{wes} P.G. Westervelt, R.S. Larson, J. Acous. Soc. Amer.
{\bf 54}, 121 (1973).

\end{thebibliography}
\end{document}